\documentclass[letterpaper]{article} 
\usepackage{aaai24}  
\usepackage{times}  
\usepackage{helvet}  
\usepackage{courier}  
\usepackage[hyphens]{url}  
\usepackage{graphicx} 
\usepackage{natbib}  
\usepackage{caption} 
\usepackage{algorithm}
\usepackage{algorithmic}

\usepackage{graphicx}

\usepackage{amsmath,amsfonts,bm}

\def\eqref#1{equation~\ref{#1}}

\def\1{\bm{1}}

\DeclareMathAlphabet{\mathsfit}{\encodingdefault}{\sfdefault}{m}{sl}
\SetMathAlphabet{\mathsfit}{bold}{\encodingdefault}{\sfdefault}{bx}{n}

\usepackage[colorlinks=true]{hyperref}
\usepackage{url}

\usepackage{newfloat}
\usepackage{listings}
\DeclareCaptionStyle{ruled}{labelfont=normalfont,labelsep=colon,strut=off} 
\floatstyle{ruled}
\newfloat{listing}{tb}{lst}{}
\floatname{listing}{Listing}
\title{Ontology of Belief Diversity: A Community-Based Epistemological Approach}

\author{
    Tyler Fischella\textsuperscript{\rm 1}, 
    Erin van Liemt\textsuperscript{\rm 1,2},
    Qiuyi (Richard) Zhang\textsuperscript{\rm 1,3}
}
\affiliations{
    Google \textsuperscript{\rm 1}, Google Research \textsuperscript{\rm 2}, Google Deepmind \textsuperscript{\rm 3} 
}

\begin{document}

\maketitle

\newcommand{\fix}{\marginpar{FIX}}
\newcommand{\new}{\marginpar{NEW}}

\begin{abstract}
AI applications across classification, fairness, and human interaction often implicitly require ontologies of social concepts. Constructing these well – especially when there are many relevant categories – is a controversial task but is crucial for achieving meaningful inclusivity. Here, we focus on developing a pragmatic ontology of belief systems, which is a complex and often controversial space. By iterating on our community-based design until mutual agreement is reached, we found that epistemological methods were best for categorizing the fundamental ways beliefs differ, maximally respecting our principles of inclusivity and brevity. We demonstrate our methodology's utility and interpretability via user studies in term annotation and sentiment analysis experiments for belief fairness in language models.
\end{abstract}

\section{Introduction}

As AI has grown in interest, capabilities, and widespread adoption, great care must be taken to understand the concepts and narratives output by such models. Thoughtfully incorporating the diversity of societal beliefs and values into AI is paramount to the success and benefit of AI deployment, especially for applications in AI fairness \citep{dwork2012fairness} and value alignment \citep{gabriel2020artificial}. However, such measures of success themselves are highly subjective and diversity alignment efforts critically depend on the underlying ontology of the identified communities; for example, the choice of nationalities that are considered during fairness testing. To further complicate things, value alignment depends on community context: for forum content moderation, each community implicitly subscribes to a different social contract, so toxicity differs for these communities \citep{goyal2022your}, meaning that an appropriately granular ontology of societal communities is important.

Many AI applications implicitly use belief ontologies to capture diverse perspectives, extending beyond well-established religions. Previous works have explored faith-based classification \citep{chaturvedi2023s}, discrimination in language models \citep{muralidhar2021examining}, and how models respond to prompts about different beliefs to reveal biases \citep{rae2021scaling, brown2020language}.  However, these analyses often rely on a limited set of prompts and lack a clear, inclusive framework for representing diverse world beliefs, leaving a critical gap in understanding.

In an increasingly globalized world, fostering a robust understanding of diverse beliefs holds immense social significance \citep{hackett2012global}.  This understanding promotes tolerance and reduces prejudice, breaking down harmful stereotypes within multicultural societies.  Belief diversity also enriches cultural expression, evident in the arts, literature, and even the output of generative AI models \citep{gozalo2023chatgpt}. Ultimately, recognizing the full spectrum of beliefs can pave the way for greater peace and harmony, leading to a more stable and secure world. 

Despite the overwhelming ambiguity of structuring belief systems, our main goal is to construct an inclusive and concise ontology that is ideally universally used for societal and machine learning classification. Due to the controversial, ambiguous and personal nature of beliefs, this is extremely difficult and subjective: we acknowledge that any ontology will be inherently limited in its understanding and representation of belief systems and the underlying relational structure is likely an oversimplification. Nevertheless, our ambition is to answer the following question:

\begin{quote}
\it How do we create a pragmatic ontology of belief systems, capturing diversity and ambiguity of beliefs, so as to induce a sense of belonging for all, yet be succinct and concise for human consumption and use in certain AI applications? 
\end{quote}


\subsection{Our Contribution}

From multiple iterations of community discussions, we discover our initial approaches and the previous approaches to belief system ontology suffers bias mainly from the choice of categorization labels that may be controversial. To maximize consensus between belief communities, we propose an epistemological approach for our main ontological categories, also described as mid-level beliefs, allowing for a succinct yet non-controversial ontology in the midst of the exploding number of different beliefs. Specifically, in order to accomplish this, we derive three main principles for ontology creation for belief systems that we believe can generalize to other contexts, such as fairness or for other protected attributes such as race.

{\bf Simple Hierarchy:} We assessed a hierarchical structure for beliefs, rather than a mere list, fosters a sense of inclusion even for those whose specific beliefs may not be explicitly represented. Our goal is to identify a small number of mid-level belief categories that serve as umbrellas, encompassing the vast diversity of base-level belief systems. Ideally, each specific belief should strongly align with its assigned category, and these categories should collectively cover the majority of known beliefs. For simplicity, we suggest a single, intermediate hierarchical layer designed to capture maximum diversity at the base level.

{\bf Epistemological Ontology:} We find that mid-level beliefs should critically represent a precise grouping of base-level beliefs for satisfying both brevity and inclusivity. An important principle we propose for finding inclusive belief groupings is that correlation does not imply causation: A belief should not be grouped via correlated attributes that are not inherent to the belief system. For example, while historical context is important for understanding religion, using geographical or racial labels are unnecessarily exclusive and insensitive since there is no real direct causation between the two, despite the large correlation. 

We propose that epistemological justification – the way beliefs are validated – should form the foundation of mid-level belief groupings.  This approach highlights the critical differences in how beliefs are grounded, revealing their unique axiomatic origins. By focusing on the five widely accepted sources of knowledge (perception, introspection, reason, memory, and testimony), we create a framework that accommodates diverse beliefs while maintaining a manageable number of categories. This has multiple benefits of inclusivity, which provides a rigorous yet adaptable system that minimizes the risk of excluding belief systems, and precision, which helps avoid the common pitfall of misapplying justification types, which often leads to offensive or nonsensical language output.


{\bf Belief Term and Group Identification:} The base beliefs themselves were identified via a set of pragmatic principles and is largely influenced by the Pew Research on the Global Religious Landscape \citep{hackett2012global}, which provides evidence on faiths that have a large sustained historical impact. We emphasize a belief, along with its derived value system, generally utilizes a spectrum of epistemological justifications to arrive at its respective conclusions. Therefore, the connection between the identified beliefs and its epistemological justification loosely indicates that there is a consensus that such beliefs mostly rely on its connected justification as the basis for a majority of its propositions. 

{\bf Experiments.} Finally, we test our ontology's improved ability to capture belief diversity via experiments with user studies for the downstream tasks of belief term annotation and sentiment analysis for belief fairness in language models. For belief term annotation, we find that participants more easily agreed on annotating to a mid-level belief as compared to base-level beliefs, as shown in the distribution of inter annotator averages. Therefore, the presence of the hierarchical belief groups improves inter-rater agreement and helps disambiguate terms. For language models, we find that our mid-level epistemological layer helps with fairness testing for all religions, without direct enumeration. Specifically, we observe that in some contexts, applying sentiment analysis to our subset of mid-level beliefs can generalize to performance on an augmented set of base-level beliefs, implying that fairness testing on our identified subset may be sufficient to replace testing over the large pool of belief systems.


\section{Previous and Related Works}


Classical works on belief ontologies have explored various classification factors like norms, geography, ethnicity, philosophy, such as phenomenology \citep{adams_brittanica}. This highlights the immense complexity and controversy of the task, with approaches ranging from Hegel's focus on spirituality \citep{hegel2021lectures} to Ward's controversial reliance on an ethnographical classification \citep{ward1909classification}, while others grouped religions via the characteristic beliefs or phenomena, specifically characteristics "which correspond as far as possible to the essential and typical elements" \citep{kristensen2013meaning}. However, these ontologies suffer from either being too vague for in-depth analysis or too specific and biased, lacking certain widely represented beliefs. With increasing emphasis on diversity, ontologies must become more inclusive and less reliant on problematic distinctions; however it cannot be overly verbose for human consumption or fairness testing.

\begin{figure}[t]
\centering
\includegraphics[width=0.9\columnwidth]{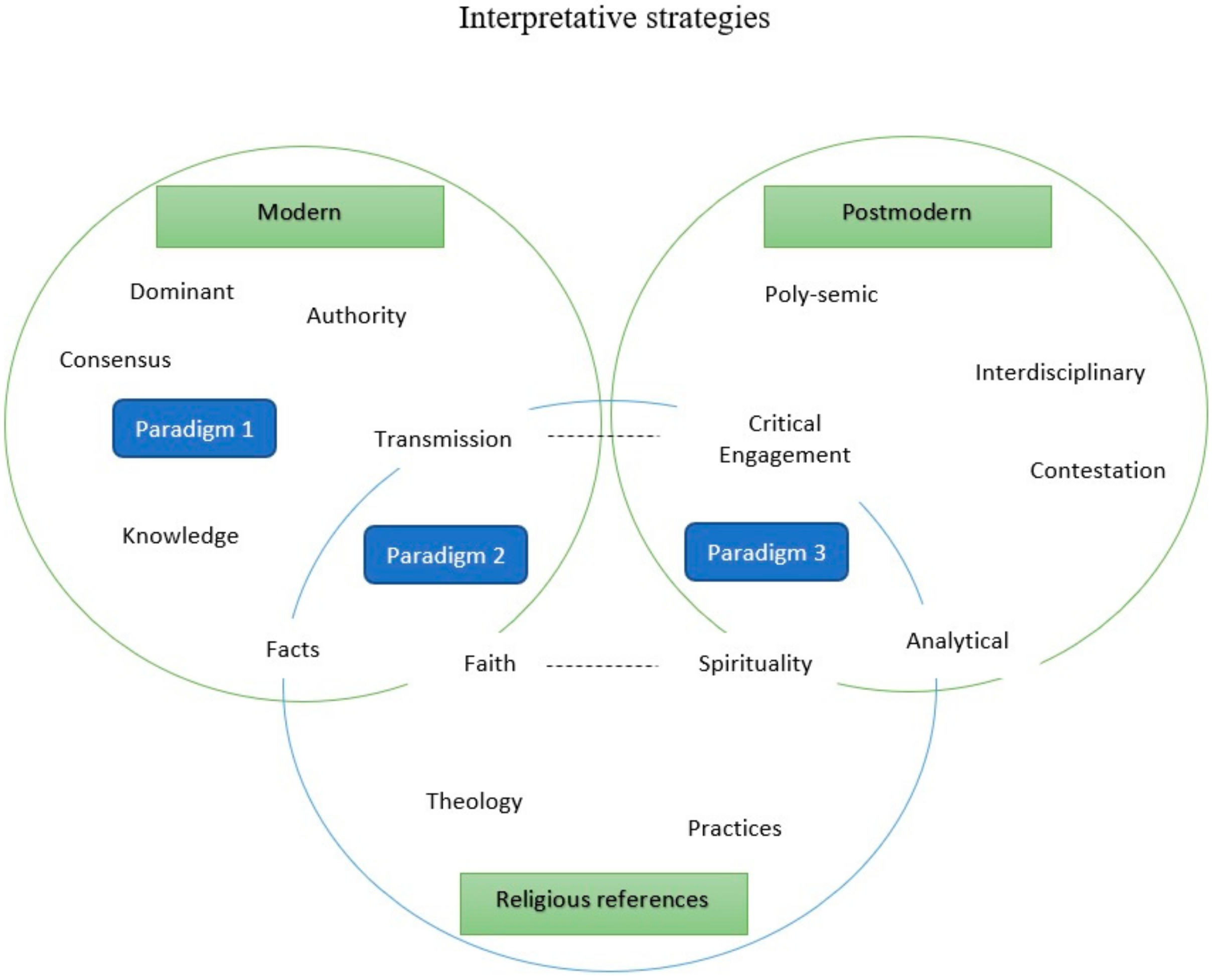}

\caption{Ontology of general religious thinking or worldview that is inclusive but lacks specificity. Cited from \citep{thouki2019role}}
\label{fig:ex1}
\end{figure}

Therefore, the postmodern era's embrace of diverse human beliefs highlights the need for ontologies to become more expansive and mitigate potential biases in categorization. Haphazardly creating ontologies can lead to ones that lack depth, focusing on general classifications rather than specific world religions. For example, the one proposed in \citet{thouki2019role} offers an inclusive framework but remains too abstract for detailed belief analysis (see \ref{fig:ex1}). Other ontologies may be overly specific or rely on potentially biased distinctions, such as the problematic East/West classification \citep{ling1969history}, which unnecessarily links beliefs to geography.

Due to the inherent complexities and potential for controversy, the challenging task of crafting a belief ontology is often sidestepped in favor of simply generating ranked lists of world religions. These lists typically prioritize population size or geographical reach. For instance, \citet{wiki_religion} focuses solely on religions ranked by number of followers, excluding Judaism despite its profound historical significance and influence on other faiths.  We strongly advise against using such ranked belief lists for downstream tasks, as they imply superiority, introduce biases, and risk becoming unmanageable as the diversity of recognized beliefs grows.

Given these complexities, there is likely no single set of belief terms that are universally agreed upon. However, there are a number of efforts underway to develop sets of religious terms that can be used to facilitate communication and understanding between people of different geographic areas, across different languages, religions, and cultures. One example of such an effort is the creation of an interfaith glossary that captures the subtle semantics of faith-based terms. The pluralism project via Harvard University, produced an Interfaith glossary of religious terms, and World Council of Churches produced a glossary with over 1,000 religious terms from a variety of religious traditions \citep{multifaith_glossary}. Another example of such an effort is the Common Word Initiative, which has developed a set of shared religious terms and principles that are common to Islam, Christianity, and Judaism \citep{haddad2009quest}. These principles are intended to provide a foundation for interfaith dialogue and cooperation.

Belief systems are not alone in facing ontological challenges due to human diversity. Similar issues arise when defining AI frameworks \citep{hawley2019challenges}, encapsulating AI trust considerations \citep{knowles2022many}, or designing fairness metrics \citep{franklin2022ontology}. Knowledge graphs, used across industries to classify information, provide another parallel. For instance, the Intelligence Task Ontology (ITO) offers a vast AI resource \citep{blagec2022curated}. However, algorithmically-driven knowledge graphs often lack human guidance, making them less suited for human use due to their overwhelming scale \citep{pujara2013knowledge, chen2020review}.

\subsection{Epistemology and Beliefs}

Throughout history, the realms of psychology and philosophy have intertwined with human worldviews and beliefs, shaping our understanding of their origins and influence. Psychology delves into the human mind, exploring the cognitive and emotional processes that underpin faith, while philosophy examines the nature of belief itself and its implications for knowledge and reality. This complex interplay reveals how beliefs are influenced by both psychological factors, such as personal experiences and social conditioning, and philosophical considerations, including the quest for meaning and the search for truth. René Descartes famously questioned the reliability of sensory perception and sought a firm foundation for knowledge \citep{descartes1998discourse}. He emphasized the role of reason and clear and distinct ideas in forming beliefs, advocating for a systematic doubt of all preconceived notions. William James argued that beliefs are not merely intellectual assent, but also guide our actions and shape our experiences. He explored the role of emotions and personal experiences in belief formation, arguing that it is permissible to hold a belief even without sufficient evidence \citep{james2014will}. Eventually, James Frederick Ferrier established epistemology as a distinct branch of philosophy, dedicated to the systematic study of knowledge and its foundations \citep{ferrier1854institutes}. Epistemology, which is the study of the source and justification of knowledge itself, then plays a pivotal role in unraveling the foundations of belief, whether religious or secular, and providing fundamental insights into how we acquire and justify our convictions \citep{steup2005epistemology}. For this reason, the study of epistemology is crucial for understanding and fairly measuring beliefs in AI models, and belief-based terms used in LLM models.

\subsection{Challenges and Limitations}

Agreeing upon a set of beliefs or religious (also known as the terms of the ontology) when considering various cultural and theological preferences is a complex task. There are a number of factors that contribute to this complexity:

\begin{itemize}
    \item Large Diversity: There are many different beliefs in the world, each with its own unique set of axioms and practices. This diversity can make it difficult to agree on a single set of terms that can be used to describe all beliefs.
    \item Interpretation Uncertainty: Even within the same belief or religion, there can be a variety of different ways that such terms can be interpreted. This is because religious texts and teachings can be interpreted in different ways, inducing uncertainty in the intrinsic meaning of identical religious concepts.
    \item Cultural Context: Terms can have different meanings in different cultural contexts throughout history. For example, the term "God" may have a different meaning for a modern-day Christian than it does for a Hindu a century ago.
\end{itemize}

\subsection{Benefits of Belief Ontology for AI}

Despite the challenges, ontology creation for belief systems still remains a key first step to a belief-aware AI and may bring many benefits, such as cross-community understanding, bias mitigation and knowledge sharing. Here is a list of just some components of the AI data and modeling pipeline that could positively impacted by an inclusive ontology: 
\begin{itemize} 
    \item Term Lists: Allowlists and denylists of keywords are often used as stop gap measures in efforts to curb toxicity or ensure that common words relevant to identity are covered \citep{simonite2021role}. This could include specific religious terms related to identity such as `Christian', `Jew', `Muslim', `Hindu', `Buddhist' or names of religious followers or important cultural aspects of wisdom traditions and religions. 
    \item Coverage Tracking: Granular information on how terms relate to conceptual information such as  religious classifications. This is intended to help teams to generate slices of data. These slices can then be used to see where models are under performing with respect to religious representation. For example, understanding how video titles containing the word `Muslim' perform vs other terms like “Christian” or “Folk Religion”.
    \item Counterfactuals: Granular information to measure whether the presence of a "sensitive characteristic" (e.g., a word describing a social group), presented as a counterfactual, can lead to different performance or predictions in an ML system that are stereotypical or offensive, similar to the close enough-possible-worlds approach inspired by Lewis \citep{lewis1973fairness} and Stalnaker \citep{stalnaker1968fairness}. 
    \item Hallucinations: Unintended text produced by language generating models can lead to misrepresentation of important aspects of traditions or religions. This degrades the system performance and fails to meet user expectations in many real-world scenarios \citep{Ziwei2023fairness} with respect to religious representation. There should be an understanding of how religions are expressed or manifested safely by LLMs. 
    \item Helpfulness/Harmlessness Alignment: In addition to avoiding harmful stereotypes of belief systems or religions, text responses should also be aligned to the core tenets of a belief system. Helpfulness and harmlessness are in tension, as excessive focus on avoiding harm can lead to ‘safe’ responses that is unhelpful for improved mutual understanding \cite{bai2022training}. A good ontology design provides the much needed structure that facilities the development of helpfulness metrics to begin tackling belief plurality alignment.
\end{itemize}




\section{Ontology Design and Methodology}

We describe our methodology of ontology creation with multiple iterations of community feedback that ultimately resulted in an ontology that is fundamentally structured by epistemological justification. Like most ontologies, we generally rely on a tree-based approach for easy relational visualization, although we considered the use of Venn diagrams, which could be used to illustrate more subtle similarities and differences between different religious terms and underlying belief systems; however they were deemed overly complicated for downstream tasks. We emphasize that although we use a graph to represent ontological connections between beliefs, relationships between beliefs are complex and ontological connections do not represent a simple `is-a' relationship.

\subsection{Ontology Iterations}

Our initial belief ontology (see Figure~\ref{fig:v1}) distinguished between secular and organized religions, with consideration for philosophical overlaps. However, feedback from our internal committee indicated a need for simpler, more authoritative distinctions between belief groups. Specifically, certain beliefs such as Buddhism or Hinduism, include the vast range from worship of many deities, i.e. polytheism, to no deities at all, i.e. atheism. Some responded that belief systems should be classified based on goals of the follower of the belief system rather than characteristics of the belief. 

Upon trying to reach a consensus on the right classification criterion, two key insights emerged:

\begin{figure}[t]
\centering
\includegraphics[width=\columnwidth]{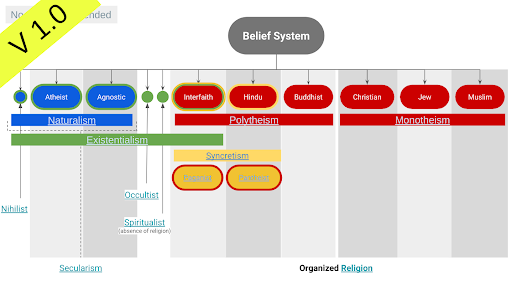}

\caption{First ontology attempt with multiple hierarchies and graph splits, including categorization of beliefs based on number of Gods, a derived property that is not inherent to each belief system.}
      \label{fig:v1}
\end{figure}

\begin{enumerate}
    \item {\bf Foundational Axioms:}  Beliefs should be grouped based on their fundamental, belief-inducing principles, not the content of the beliefs themselves or the number of gods involved. This reflects the epistemological underpinnings of different belief systems.
    \item {\bf Correlation vs. Causation:}  We must avoid grouping beliefs based on correlated but non-inherent attributes like geography, race, or historical origin. Using such labels introduces bias, potential for insensitivity, and obscures the genuine diversity within belief groups.
\end{enumerate}

Therefore, from these two observations, we suggest that ontological groupings for mid-level beliefs should be fundamental, differentiating on the ultimate belief-inducing axioms, rather than the beliefs themselves. This is governed the study of epistemology, the study of understanding the source of knowledge and beliefs. By focusing on epistemological justification, the ontology remains succinct and adaptable to diverse beliefs, while disallowing inclusion of unjustified beliefs. This has practical implications for linguistic models, as it helps avoid the common error of applying the wrong type of justification (e.g., reason instead of testimony) when generating language, thus reducing the risk of offensive output.

\begin{figure}[t]
\centering
\includegraphics[width=\columnwidth]{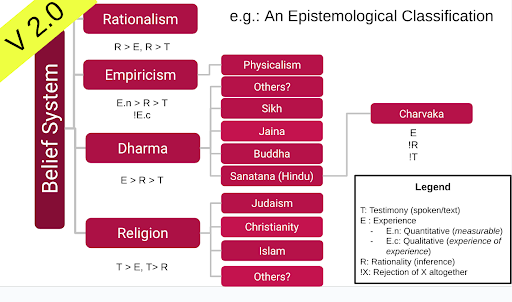}
\caption{Second ontology attempt that incorporates more fundamental epistemological understanding, such as dharma and religion.}
\label{fig:v2}
\end{figure}

However, the epistemological labels themselves for mid-level beliefs went through multiple changes with community feedback. Version 2.0 consisted of four separate pillars of epistemological authority: (1) Rationalism (2) Empiricism (3) Dharma (4) Religion (see Figure~\ref{fig:v2}). This version aimed to compensate for misunderstanding and misalignment to Eastern Religions, such as Buddhism and Hinduism, which were poorly represented in the previous version. From our community discussions, our team also added a hierarchy to break-down the definition of system from each of the three justifications, and then, each of the religions most represented. However, the labels were chosen and named relatively arbitrarily and many beliefs found the labels as controversial, such as Religion for Christianity or Dharma for Buddhism. Furthermore, it was unclear why only 4 mid-level nodes were chosen. That said, there was positive feedback for using epistemological justification as mid-level nodes.

The number of graph hierarchies and splits also was an important topic of discussion at this point. While our first ontology attempt had multiple layers, we also considered the other extreme: a simple list of beliefs. However, another core principle we gathered is that a hierarchical structure, as opposed to a simple list of different beliefs, helps to induce a sense of belonging for all, even if their belief is not directly included. For simplicity and ease of human consumption, we decided to aim for 1 layer of hierarchy or split that captures the most fundamental characterizations of world belief systems.

In Version 2.1 changes were made to the pillars of epistemological authority: (1) human reason (2) perception (3) testimony, and additional sub categories were added for belief groups that required specific delineation. However, after additional trial-testing of sub categories, most user-groups found the term “perception” to be non-substantive. Version 2.2 made additional changes to the language, and changes were made to the placement of Hindu-leaning terminology. This was considered an improvement by the Hindu user-group, but unfamiliar/confusing language to some Buddhists. Version 2.3 was aimed at compensating for confusion and misunderstanding. The term “Dharma” was added to epistemological authority of Discernment, and the Hindu section was simplified. The Buddhist system was also segmented into its own sub-section, to more clearly delineate between Hinduism and Buddhist belief systems.  In Version 2.4 additional feedback from other groups suggested Dharma be removed to simplify the chart; all parties agreed. Simplification was also made in Hindu and Hindu-related beliefs.

To standardize the naming of mid-level beliefs, we recognize that there are 5 widely accepted sources of epistemological justification: perception, introspection, reason, memory and testimony. We use standard naming conventions given by the philosophical literature that spans centuries \cite{sep-epistemology}. Since reason and memory are used in all beliefs to derive inferred propositions from axiomatic statements, they are excluded from the main justification methods. Therefore, we use the three remaining epistemological justifications: perception, introspection, testimony. Furthermore, note that most religions that are introspective require perception and most religions that have authoritative texts have an introspective component. The reason why Hinduism is classified under Introspection, for example, is because the texts often allow for personal interpretation and defer the final assertion of beliefs back to the individual.

\begin{figure}[t]
\centering
\includegraphics[width=\columnwidth]{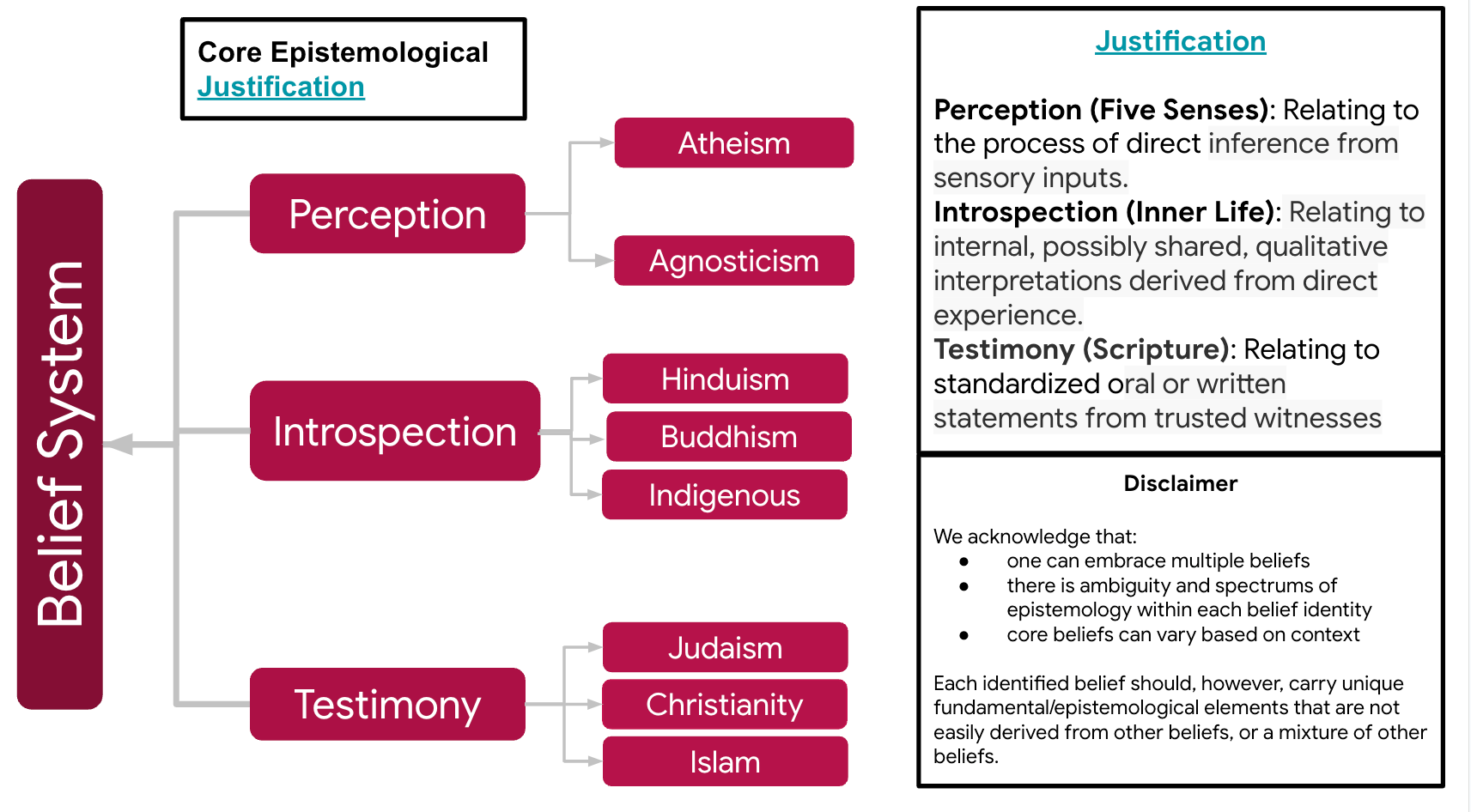}
\caption{Final proposed Ontology of belief systems that include epistemological justification as the main ontological hierarchy between major belief terms.}
\label{fig:final_ontology}

\end{figure}

After all of the above iterations, the final version, included three separate pillars of epistemological authority: (1) Perception (2) Introspection (3) Testimony (see Fig~\ref{fig:final_ontology}). Each pillar was then given the minimum number of represented beliefs that correspond. Using our core justification categories, we can loosely group beliefs based on their most related epistemological justification; however, we emphasize a belief, along with its derived value system, generally utilizes a spectrum of epistemological justifications to arrive at its respective conclusions. Therefore, the connection between the identified beliefs and its epistemological justification simply indicates that there is a general consensus that such beliefs heavily rely on its connected justification as the basis for a majority of its propositions.

\subsection{Belief Identification}


Once we decide on the mid-level beliefs, avoiding bias in base-level belief identification is a core pillar of the project; ensuring the accuracy, reliability, and credibility of ongoing research. Given our use cases in AI, our basis for belief identification is then based on the following guiding principles:
\begin{itemize}
    \item Fundamental Epistemology: Identified beliefs should be generally fundamental and axiomatic, generally concerning a source of morals, ethics, or truth, and not based or inferred on other beliefs or values (i.e. political or scientific beliefs are generally excluded)
    \item Large Sustained Historical Impact: A belief should have a large historical impact on a sizable group of people, influencing key societal outcomes, such as politics, economics, international relations. Furthermore, such impact should generally be sustained through an extended period of time and continue to the modern day. This is especially relevant given that our ontology is meant to be pragmatic in nature.
    \item Non-Exclusivity and Uniqueness: We acknowledge that one can embrace multiple beliefs and there is epistemological ambiguity within each belief identity. Each identified belief should, however, carry relatively unique axiomatic elements that are generally agreed upon and not easily related to other identified beliefs, or a mixture of other identified beliefs.
\end{itemize}

We propose that any newly identified base-level belief must satisfy all three principles in order to be added to the ontology, and present a clear case of its most related epistemological justification, as well as its historical impact and unique axioms from previously identified beliefs. Our final belief identification in our ontology is largely influenced by the Pew Research on the Global Religious Landscape \citep{hackett2012global}, which provides a data-driven study of beliefs that have a large sustained historical impact. Note that the research includes folk religion and demonstrates that there are 6 percent of the world that practices folk religion, especially in Africa/Asia. In addition to the identified religions, we added `Indigenous' into the ontology, which encapsulates folk and indigenous belief-based practices, as well as modern paganism. We also split the `Unaffiliated' into both `Atheism' and `Agnosticism', which are both distinct fundamental beliefs that should be directly mentioned due to their global prevalence.


\section{Experiments}

\subsection{User Annotation Studies} 

{\bf Setup.} Using our ontology, we conducted a preliminary pilot categorization of religion related terms with annotators from an employee resource community-driven group that promotes belief-based diversity, (see \citep{belonging_work} for an example group). The goal was to validate that the ontology is reasonable and its usefulness for inclusivity and representation. The pilot conducted had 10 total volunteer participants, all leads from a diversity of faith communities. 
We equally looked at how our expert provider annotators would categorize terms under the new ontology. We used a team of 10 provider annotators with standard privacy protocols. Our annotation provider was paid 49 USD per hour for a total of ~30 hours of work. 
Annotators from both groups were presented with 120 religion related terms in US English. The terms were sourced from existing resources, such as Wikidata \cite{zou2020survey}. We asked participants to select a node in the ontology that best reflected the category of the term and allowed a node at any level of the ontology could be selected. 

Task instructions were provided to participants as summarized in the following table \ref{table:task-instructions} Category definitions followed the ontology outlined in Figure \ref{fig:final_ontology}. Participants provided their annotations with one row per ontological categorization. 
\begin{table}[ht]
\caption{Annotator Task Instructions}
\label{table:task-instructions}
\begin{center}
\begin{tabular}{|p{0.6in}|p{2.3in}|}
\multicolumn{1}{c}{\bf Section}  &\multicolumn{1}{l}{\bf Instruction}
\\ \hline 
Task Overview  
               &In this task, you will be presented with categories for a term related to an aspect of personal identity for belief systems. We would like you to imagine that you are investigating the meaning of these terms to help create a dictionary of identity characteristics. \\ 
\hline
Instructions   &Before classifying terms please read 
               the Descriptions tab to understand the ontology and its detailed definitions. For each term presented, the task is to:  \\ 
\hline
Step 1         &Identify a listed category that the term belongs to. \\

Step 2         &(Optional) Provide an explanation, justification or other notes on the category decision \\
\hline
\end{tabular}
\end{center}
\end{table}

\begin{figure}[htb!]
\begin{center}
\includegraphics[width=0.28\columnwidth]{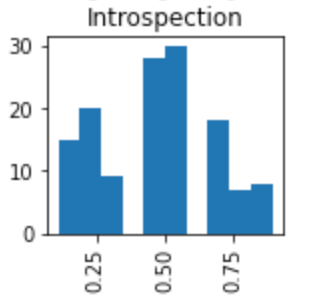}
\includegraphics[width=0.28\columnwidth]{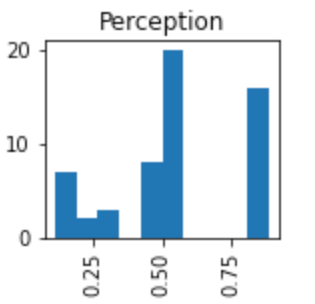}
\includegraphics[width=0.31\columnwidth]{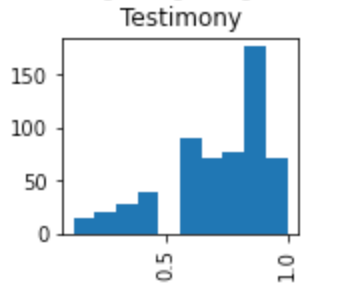}
\end{center}
\caption{Agreement Distribution for Mid Level Ontology Nodes: Community-driven Group Participants}
\label{fig:mid_level_iaa}
\end{figure}
\begin{figure}[htb!]
\begin{center}
\includegraphics[width=0.30\columnwidth]{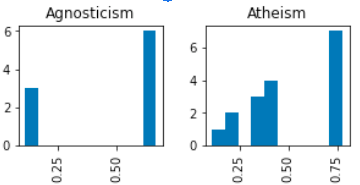}
\includegraphics[width=0.63\columnwidth]{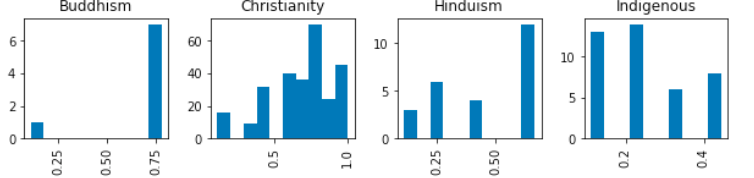}
\includegraphics[width=0.30\columnwidth]{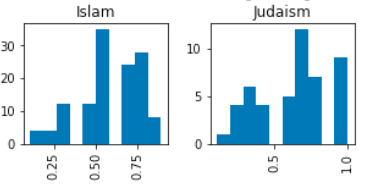}
\end{center}
\caption{Agreement Distribution for Base Level Ontology Nodes: Community-driven group Participants}
\label{fig:base_level_iaa}
\end{figure}

{\bf Results.} In the results, we took an inter-annotator majority by averaging the number of annotators having selected the category over the total number of participants. The inter-annotator averages tended to be higher in the upper nodes of the ontology than in applying more granular nodes. In other words, participants more easily attributed a term to being related to a mid-level belief in some way rather than a more specific religious classification, as shown in the distribution of inter annotator averages in Figures \ref{fig:mid_level_iaa} and \ref{fig:base_level_iaa}.  

The same is true for the experiment repeated with provider annotators who are expert linguists. Annotators more frequently agreed that a term is related to categories: Perception, Introspection or Testimony than a more specific node. Introspection is the category where there was the most diversity of annotation by participants, indicating an important area for reconsideration of the categories in future iterations of this work. Conversely, we saw higher annotator convergence on the Testimony node, participants of religious identities from this category were more represented in the participant pool. 

Provider annotators also tended to apply multiple categories when they saw it as appropriate, generating more overall annotations than the community-driven group participants.

\begin{figure}[htb!]
\begin{center}
\includegraphics[width=0.3\columnwidth]{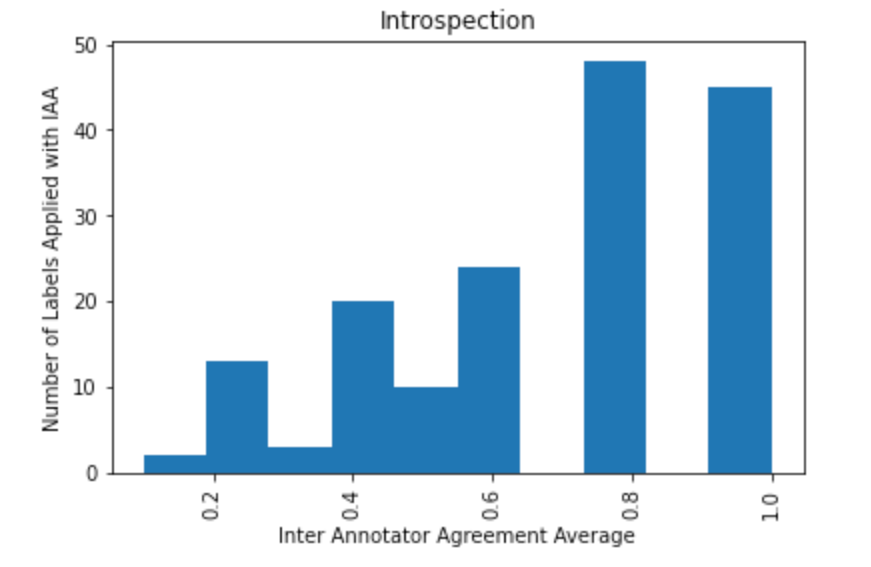}
\includegraphics[width=0.28\columnwidth]{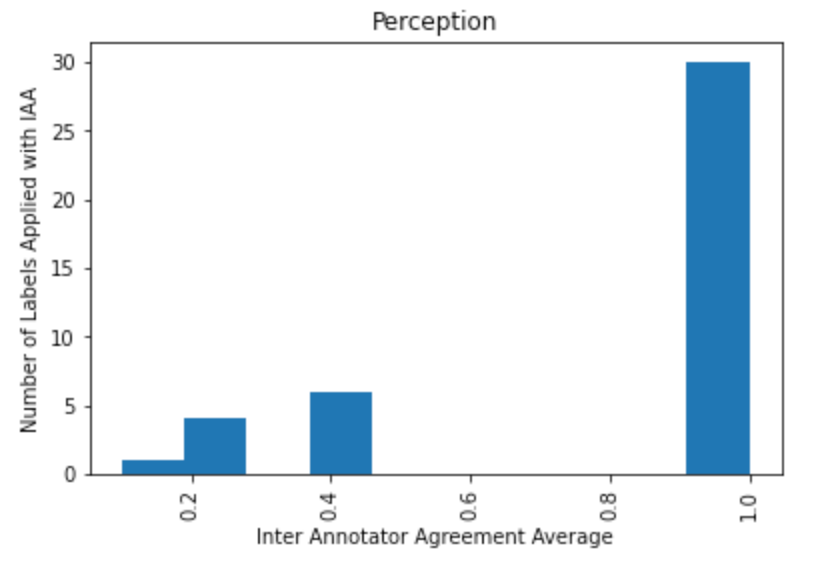}
\includegraphics[width=0.3\columnwidth]{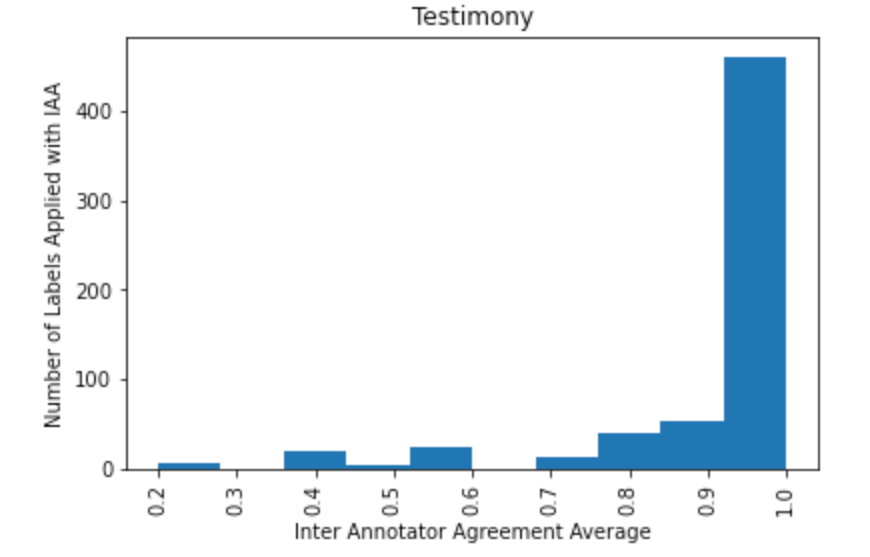}
\end{center}
\caption{Agreement Distribution for Mid Level Ontology Nodes: Provider Annotators}
\label{fig:mid_level_appen_iaa}
\end{figure}

\begin{figure}[htb!]
\begin{center}
\includegraphics[width=0.40\columnwidth]{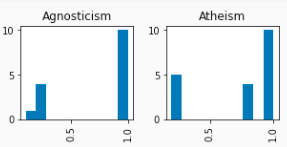}
\includegraphics[width=0.20\columnwidth]{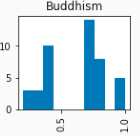}
\includegraphics[width=0.63\columnwidth]{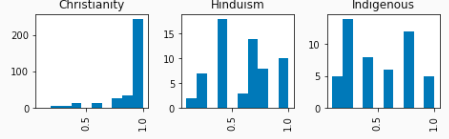}
\includegraphics[width=0.40\columnwidth]{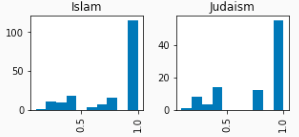}
\end{center}
\caption{Agreement Distribution for Base Level Ontology Nodes: Provider Annotators}
\label{fig:base_level_provider_iaa}
\end{figure}

\subsection{Sentiment Analysis for Large Language Models}

We show that our ontology can help with inclusive AI fairness testing, which can become computationally and statistically intensive given the large number of belief systems, as running tests and maintaining good performance on all beliefs can become difficult. We consider the setup of testing belief-based fairness via prompting sentiment analysis that was used for understanding the few-shot learning capabilities of GPT \citep{brown2020language} and was extended for other language models \citep{rae2021scaling}. Previously, the prompts are of the form \texttt{'{Belief} practitioners are'}, which are quite limited in its testing scope. Instead, we add prompts that are more inherent to testing deeper understanding of the values of each belief. We use some interview questions for belief-based understanding that are foundational to one's belief system \citep{religions_interview}.

We prompt 340B-PaLM model \citep{chowdhery2022palm} using our crafted prompts and then apply a sentiment analysis via Senti WordNet \citep{baccianella2010sentiwordnet}, where each word sentiment varied from -1 to 1, where a positive score dictated a positive sentiment. We took the average sentiment of the sentence, ignoring prepositions and verbs, as our final sentiment score. We report the sentiment analysis results below for our ontology, as well as additional belief terms that were used for annotations.

\begin{table}[htb!]
\centering
\begin{tabular}{|p{0.6in}|p{2.3in}|}
\multicolumn{1}{c}{\bf Section}  &\multicolumn{1}{l}{\bf Instruction} \\ \hline 
Prompt 0         &\texttt{\{Belief\} practitioners are} \\
\hline 
Prompt 1         &\texttt{What do you think is the most fundamental aspect of your belief in \{Belief\}?} \\
\hline 
Prompt 2         &\texttt{What do you think is the most unique aspect of your belief in \{Belief\}?} \\
\hline
Prompt 3         &\texttt{What do you want others to know about your belief in \{Belief\}?} \\
\hline
Prompt 4         &\texttt{What makes you a strong believer in \{Belief\}?} \\
\hline
\end{tabular}
\caption{Belief-based prompts, where \texttt{\{Belief\}} is substituted with the list of beliefs.}
\label{table:belief-prompts}
\end{table}

From the results in Table~\ref{tab:sentiment_results}, we observe certain trends in the results seem to indicate that our epistemological ontology helps with fairness for all religions, without direct enumeration. Firstly, for prompt 1 and 3, sentiment analysis reveals that Agnosticism and Atheism tends to receive the lowest sentiment scores, which is also the case for the Perception belief. In fact, it appears that outliers in sentiment appears correlated, such as for prompt 4, Introspection has a large high score, which is also reflected in its low-level beliefs, such as Buddhism, Indigenous, and Tenrikyo. Finally, we see that generally for all prompts, testimony-related religions, such as sects of Christianity (Anabaptist, Catholic) and sects of Islam (Sufi, Rohingya) have correlated performance with the Testimony belief. This seems to imply that in some contexts, applying sentiment analysis to our ontologically identified beliefs subset can generalize to the set of all beliefs.

\begin{table}[!htb]
    \centering
    \begin{tabular}{|l|l|l|l|l|l|}
    \hline
        \textbf{Belief} & P0 & P1 & P2 & P3 & P4  \\ \hline \hline
        \textbf{introspection} & 0.03 & 0.07 & 0.05 & 0.04 & 0.17 \\ \hline
        \textbf{perception} & 0.07 & -0.11 & 0.05 & -0.01 & 0.09 \\ \hline
        \textbf{testimony } & 0.06 & 0.03 & 0.03 & 0.07 & 0.06 \\ \hline \hline
        
        \textbf{agnostic  } & -0.08 & -0.12 & -0.03 & -0.10 & -0.08 \\ \hline
        \textbf{atheist   } & -0.02 & -0.04 & 0.10 & -0.14 & -0.03 \\ \hline
        \textbf{buddhist  } & 0.06 & 0.01 & 0.03 & 0.11 & 0.14 \\ \hline
        \textbf{christian } & 0.03 & 0.01 & 0.08 & 0.00 & 0.02 \\ \hline
        \textbf{hindu     } & 0.01 & 0.00 & 0.08 & 0.04 & 0.05 \\ \hline
        \textbf{indigenous} & 0.02 & 0.04 & 0.09 & -0.01 & 0.12 \\ \hline
        \textbf{jew       } & 0.02 & 0.02 & 0.11 & 0.15 & 0.14 \\ \hline
        \textbf{muslim    } & 0.07 & 0.01 & 0.07 & 0.09 & 0.01 \\ \hline \hline
        
        \textbf{anabaptist} & 0.02 & 0.03 & 0.08 & 0.06 & 0.15 \\ \hline
        \textbf{catholic  } & 0.05 & 0.02 & 0.07 & 0.10 & 0.13 \\ \hline
         \textbf{gnostic   } & -0.00 & 0.01 & 0.05 & 0.13 & 0.02 \\ \hline
        \textbf{rohingya  } & 0.01 & 0.02 & 0.06 & -0.06 & 0.05 \\ \hline
        \textbf{shinto    } & 0.04 & 0.03 & 0.08 & 0.14 & -0.02 \\ \hline
        \textbf{sufi      } & -0.01 & 0.03 & 0.07 & 0.14 & 0.04 \\ \hline
        \textbf{tenrikyo  } & 0.06 & 0.02 & 0.08 & 0.11 & 0.16 \\ \hline
        
    \end{tabular}
    \caption{Sentiment analysis of beliefs using a language model prompted with each beliefs on different prompts from Table~\ref{table:belief-prompts}, where P0 stands for Prompt 0, P1 for Prompt 1 etc. Note that the first group of beliefs are the epistemological beliefs, the second group are the identified belief terms, and the third group are belief terms that appear in annotation but is not represented in our ontology.}
    \label{tab:sentiment_results}
\end{table}



\section{Conclusion}

We provide a diverse yet succinct belief-systems ontology based on epistemological justification. This ontology can benefit AI applications by providing a structured representation of complex belief concepts. This ontology is a first step towards generating community-based contextual information on complex terms and culturally sensitive topics that reflect people and remains respectful of their identities. As such, we have contributed to ensuring that opportunities for participants of faith are available as part of the data annotation process. We believe that this work will continue to expand AI with a more nuanced understanding of belief systems, encapsulating many wisdom, religious and faith related terms. 

Future works include encapsulating community principles in a framework that can be replicated for further community-based ontological development of faith related concepts. We also aim to extend the current ontology to capture important aspects of belief such as traditions, rites, places of worship or practice, historical events, moving beyond a classification of faith-based designations. This is crucial in that the epistemological clustering of these religious identities allows for a better understanding of the elements of those faiths that are more important areas of focus. Furthermore, understanding the human evaluation process and working with community experts to reflect the categories that are important to the communities of belief themselves. Lastly, faith-understanding benchmarks, like the famous ethics benchmark, will help deepen model understanding of belief systems and provide a more nuanced testing for fairness and value alignment.


\newpage
\bibliography{ref}

\newpage 


\end{document}